\documentclass[aps,twocolumn,showpacs,preprintnumbers,amsmath,amssymb,superscriptaddress,floatfix,nofootinbib]{revtex4}

\usepackage{graphicx}
\usepackage{epsfig}
\usepackage{epstopdf}
\usepackage{hyperref}
\usepackage{amsmath}
\usepackage{amsfonts}
\usepackage{amssymb}
\usepackage{color}

\newcommand{\Slash}[1]{\ooalign{\hfil/\hfil\crcr$#1$}}

\begin{document}

\title{Role of the $N^*(1535)$ in the $\Lambda^+_c \to \bar{K}^0 \eta p$ decay}

\author{Ju-Jun Xie}
\affiliation{Institute of Modern Physics, Chinese Academy of
Sciences, Lanzhou 730000, China}

\author{Li-Sheng Geng} \email{lisheng.geng@buaa.edu.cn}
\affiliation{School of Physics and Nuclear Energy Engineering and International Research Center for Nuclei and Particles in the Cosmos and Beijing Key Laboratory of Advanced Nuclear Materials and Physics, Beihang University, Beijing 100191, China}

\date{\today}

\begin{abstract}

The nonleptonic weak decay of $\Lambda^+_c \to \bar{K}^0 \eta p$ is
analyzed from the viewpoint of probing the $N^*(1535)$ resonance,
which has a big decay branching ratio to $\eta N$. Up to an
arbitrary normalization, the invariant mass distribution of $\eta p$
is calculated with both the chiral unitary approach and an effective
Lagrangian model. Within the chiral unitary approach, the
$N^*(1535)$ resonance is dynamically generated from the final state
interaction of mesons and baryons in the strangeness zero
sector. For the effective Lagrangian model, we take a Breit-Wigner
formula to describe the distribution of the $N^*(1535)$ resonance. It is
found that the behavior of the $N^*(1535)$ resonance in the
$\Lambda^+_c \to \bar{K}^0 N^*(1535) \to \bar{K}^0 \eta p$ decay
within the two approaches is different. The proposed $\Lambda^+_c$
decay mechanism can provide valuable information on the properties of the
$N^*(1535)$ and can in principle be tested by facilities such as
BEPC II and SuperKEKB.

\end{abstract}

\pacs{13.75.Jz, 14.20.-c, 11.30.Rd}

\maketitle

\section{Introduction}

Understanding the nature of the $N^*(1535)$ with spin parity $J^P =
1/2^-$ has always been one of the most challenging topics in hadron
physics~\cite{Klempt:2007cp,Crede:2013sze}. In classical constituent
quark models, the $N^*(1535)$ is mainly composed of three valence
quarks, and its mass should be lower than the radial excitation, the
$N^*(1440)$, with $J^P =
1/2^+$~\cite{Capstick:2000qj,Olive:2016xmw}. This is the
long-standing mass reverse problem for the lowest spatial excited
nucleon states. Another peculiar property of the $N^*(1535)$ is that
it couples strongly to the channels with strangeness, such as $\eta
N$ and $K \Lambda$, which is also difficult to understand in the
naive  constituent quark models.

Renouncing the picture of baryons as three-quark bound
states, a different point of view consists in describing meson-baryon
scattering reactions by taking mesons and baryons as the relevant
degrees of freedom at low energies. Then, baryon excited states
manifest themselves as poles of the meson-baryon scattering
amplitude in a certain Riemann sheet in the complex energy plane.
For example, the unitary extensions of chiral perturbation theory
have brought new light to studies of baryon resonances from
meson-baryon interactions~\cite{Kaiser:1995cy,Kaiser:1996js}. In the
chiral unitary coupled-channels approach it was found that the
$N^*(1535)$ resonance is dynamically generated as a meson-baryon
state with its mass, width, and branching ratios in fair agreement
with
experiments~\cite{Nieves:2001wt,Inoue:2001ip,Bruns:2010sv,Nieves:2011gb,Gamermann:2011mq,Khemchandani:2013nma,Garzon:2014ida}.
The numerical results obtained in those studies differ to some extent, but it was found that the $N^*(1535)$ resonance couples strongly to the
$\eta N$ channel. Furthermore,  it couples  more strongly to $K \Sigma$ and $K \Lambda$
than to $\pi N$~\cite{Inoue:2001ip,Bruns:2010sv,Nieves:2011gb,Gamermann:2011mq,Khemchandani:2013nma,Garzon:2014ida}.

In the phenomenological studies, besides the large coupling of
the $N^*(1535)$ to $\eta N$, a large value of the coupling of
the $N^*(1535)$ to $K \Lambda$  is deduced in
Refs.~\cite{Liu:2005pm,Geng:2008cv,Mart:2013ida} by a simultaneous
fit to the BES data on $J/\psi \to p\bar{p} \eta$, $pK^- \bar{\Lambda} +
\bar{p} K^+ \Lambda$, the COSY data on $pp \to p K^+\Lambda$, and the
CLAS data on $\gamma p \to K^+ \Lambda$ reaction. There is also
evidence for a large  coupling of the $N^*(1535)$ to $\eta' N$ from the
analysis of the $\gamma p \to p \eta'$
reaction~\cite{Dugger:2005my} and $pp \to pp \eta'$
reaction~\cite{Cao:2008st}, and a large  coupling of the $N^*(1535)$ to $\phi N$
from the $\pi^- p \to n \phi$, $pp \to pp \phi$ and $pn \to d
\phi$ reactions~\cite{Xie:2007qt,Doring:2008sv,Cao:2009ea}.

The above-mentioned strange decay properties of the $N^*(1535)$
resonance can be easily understood if it contains large five-quark
components ~\cite{Liu:2005pm,Helminen:2000jb,Zou:2007mk}.
Within the pentaquark picture, the $N^*(1535)$ resonance could be
the lowest $L=1$ orbital excited $uud$ state with a large admixture
of $[ud][us]\bar{s}$ pentaquark component having $[ud]$, $[us]$,
and $\bar{s}$ in the ground state. This makes the $N^*(1535)$
heavier than the $N^*(1440)$ and also gives a natural explanation of its
larger couplings to the channels with strangeness~\cite{Zou:2010tc}.

Recently, it has been shown that the nonleptonic weak decays of
charmed hadrons provide a useful platform to study hadronic resonances,
some of which are subjects of intense debate about their
nature~\cite{Crede:2008vw,Chen:2016qju}. For instance, the
$\Lambda^+_c \to \pi^+ MB$ weak decays are studied in
Ref.~\cite{Miyahara:2015cja} from the viewpoint of probing the
$\Lambda(1405)$ and $\Lambda(1670)$ resonances, where $M$ and $B$
stand for mesons and baryons. In Ref.~\cite{Hyodo:2011js}, the $\pi
\Sigma$ mass distribution was studied in the $\Lambda^+_c \to \pi^+
\pi \Sigma$ decays with the aim of extracting the $\pi \Sigma$
scattering lengths. In Ref.~\cite{Xie:2016evi}, the $a_0(980)$ and
$\Lambda(1670)$ states are investigated in the $\Lambda^+_c \to
\pi^+ \eta \Lambda$ decay taking into account the $ \pi^+ \eta $ and
$\eta \Lambda$ final state interactions. The pure $I = 1$ nature of
the $\pi^+ \eta$ channel is particularly beneficial to the study of
the $a_0(980)$ state. The role of the $\Sigma^*(1380)$ state with
$J^P = 1/2^-$ in the $\Lambda^+_c \to \eta \pi^+ \Lambda$
decay is also studied in Ref.~\cite{Xie:2017xwx} where the
color-suppressed $W$-exchange diagram is considered for the
production of the $\Sigma^*(1385)$ with $J^P = 3/2^+$. In
Ref.~\cite{Lu:2016ogy} the role of the exclusive $\Lambda^+_c$
decays into a neutron in testing the flavor symmetry and final state
interaction was investigated. It was shown that the three body
nonleptonic decays are of great interest to explore the final state
interactions in $\Lambda_c^+$ decays.

Along this line, in the present work, we study the role of the
$N^*(1535)$ resonance in the $\Lambda^+_c \to \bar{K}^0 \eta p$
decay by taking the advantage of the strong coupling of the $N^*(1535)$
to the $\eta N$ channel and its large $uud s\bar{s}$ component. We
calculate the invariant $\eta p$ mass distribution within the
chiral unitary approach and an effective Lagrangian model. First, we
follow the same approach used in Ref.~\cite{Miyahara:2015cja} to
study the $\Lambda^+_c \to \pi^+ MB$ decays, but with the
hadronization of the $uud$ rather than the $sud$ cluster to get the final
$\eta p$ and from the
$s\bar{d}$ pair to get the $\bar{K}^0$. In this respect, the $N^*(1535)$ resonance is
dynamically generated from the final state interaction of
mesons and baryons in the $I = 1/2$ sector where we have
assumed that the $ud$ di-quark with $I =0$ in the $\Lambda^+_c$ is a spectator. Second, we study the $\Lambda^+_c \to \bar{K}^0
N^*(1535) \to \bar{K}^0 \eta p$ decay at the hadron level by taking a
Breit-Wigner formula to describe the distribution of the $N^*(1535)$
resonance within the effective Lagrangian model. The contributions
from other low-lying $N^*$ and $\Sigma^*$ resonances are discussed.
Fortunately, it is found that these contributions may not affect much the
results obtained here.

This article is organized as follows. In Sec.~\ref{Sec:Formalism},
we present the theoretical formalism of the decay of $\Lambda^+_c
\to \bar{K}^0 \eta p$, explaining in detail the hadronization and
final state interactions of the $\eta p$ pair. Numerical results and
discussions are presented in Sec.~\ref{Sec:Results}, followed by a
short summary in the last section.

\section{Formalism} \label{Sec:Formalism}

\begin{figure}[t]
\begin{center}
\includegraphics[scale=0.9]{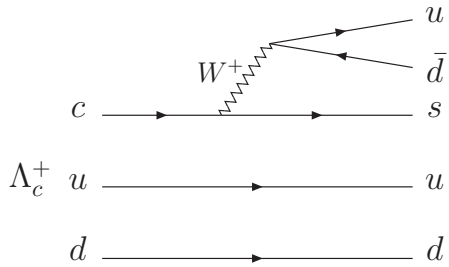}
\caption{Dominant diagram at the quark level for the charm quark
in the $\Lambda^+_c$ decaying into a $u \bar{d}$ pair and a strange
quark. The solid lines and the wiggly line stand for the quarks and
the $W^+$ boson, respectively.} \label{Fig:feynd}
\end{center}
\end{figure}

As shown in
Refs.~\cite{Miyahara:2015cja,Xie:2016evi,Miyahara:2016yyh}, a
Cabibbo allowed mechanism for the $\Lambda^+_c$ decay is that the charmed
quark in $\Lambda^+_c$ turns into a strange quark with a $u \bar d$
pair by the weak interaction as shown in Fig.~\ref{Fig:feynd}.

In addition to the $c$ quark decay process described above, in
principle one can also have  contributions from internal
$W$-exchange ($c+d \to s + u$) diagrams. As discussed in
Refs.~\cite{Miyahara:2015cja,Xie:2014tma,Xie:2016evi,Miyahara:2016yyh},
these contributions are smaller than the $c$ decay process.
Furthermore, including such contributions, the decay amplitudes
would become more complex due to additional parameters from the weak
interaction, and we can not determine or constrain these parameters at present.
Hence, we will leave these contributions to future studies  when
more experimental data become available.

\subsection{The $N^*(1535)$ as a dynamically generated state from meson-baryon scattering}

\begin{figure}[t]
\begin{center}
\includegraphics[scale=0.9]{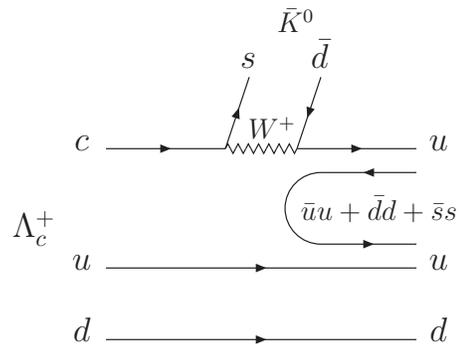}
\caption{Quark level diagram for the $\Lambda^+_c \to \bar{K}^0 MB$
decay with the $\bar{K}^0$ emission from the $s \bar d$ pair.}
\label{Fig:mbproduction}
\end{center}
\end{figure}

We first discuss the decay of $\Lambda^+_c$ to produce the
$\bar{K}^0$ from the $s \bar d$ pair and the insertion of a new
$\bar q q$ pair with the quantum numbers of the vacuum, $\bar u u +
\bar d d + \bar s s$, to construct the intermediate meson-baryon
state $M B$ from the $u ud$ cluster with the assumption that the $u$
and $d$ quarks in the $\Lambda^+_c$ are spectators in the weak decay corresponding to the mechanism of Fig.~\ref{Fig:mbproduction}. Thus, after the hadronization these $u$ and $d$
quarks in the $\Lambda^+_c$ are part of the baryon, and the
$u$ quark originated from the weak decay forms the meson.
Furthermore, the $uud$ cluster with strangeness zero is combined
into a pure $I = 1/2$ state
\begin{eqnarray}
\frac{1}{\sqrt{2}} |u(ud - du)\rangle.
\end{eqnarray}

Following the procedure of
Refs.~\cite{Miyahara:2015cja,Xie:2016evi,Miyahara:2016yyh,Roca:2015tea,Feijoo:2015cca},
one can straightforwardly obtain the meson-baryon states after the
$\bar q q$ pair production as~

\begin{eqnarray}
|MB\rangle =  \frac{\sqrt{3}}{3} |\eta p\rangle + \frac{\sqrt{2}}{2}
|\pi^0 p \rangle + |\pi^+ n\rangle - \frac{\sqrt{6}}{3} |K^+ \Lambda
\rangle, \label{Eq:mbproduction}
\end{eqnarray}
where we have omitted the $\eta' p$ term because of its large mass
threshold compared to other channels that we considered.

\begin{figure*}[htbp]
\begin{center}
\includegraphics[scale=0.8]{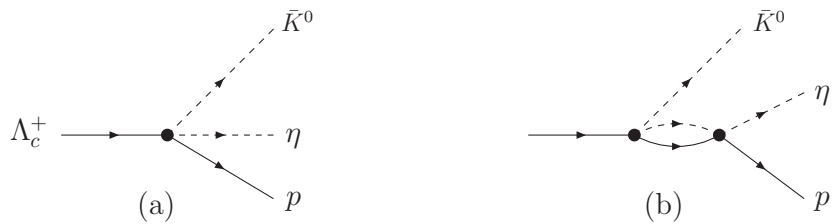}
\caption{Diagrams for the $\Lambda^+_c \to \bar{K}^0 \eta p$ decay:
(a) direct $\bar{K}^0 \eta p$ vertex at tree level, (b) final-state
interaction of the $\eta p$.} \label{Fig:mbfsi}
\end{center}
\end{figure*}

After the production of a meson-baryon pair, the final-state
interaction between them takes place, which can
be parameterized by the re-scattering shown in Fig.~\ref{Fig:mbfsi}
at the hadronic level for the $\Lambda^+_c \to \bar{K}^0 \eta p$
decay. The final-state interaction of $MB$, in $I = 1/2$, leads to the dynamical generation of the $N^*(1535)$
resonance~\cite{Inoue:2001ip,Doring:2005bx}. In
Fig.~\ref{Fig:mbfsi}, we also show the tree level diagram for the
$\Lambda^+_c \to \bar{K}^0 \eta p$ decay.

According to Eq.~\eqref{Eq:mbproduction}, we can write down the
$\Lambda^+_c \to \bar{K}^0 \eta p$ decay amplitude of
Fig.~\ref{Fig:mbfsi} as~\cite{Oller:1997yg},
\begin{eqnarray}
T^{MB} &=& V_P \Big( \frac{\sqrt{3}}{3} + \frac{\sqrt{3}}{3} G_{\eta
p} (M_{\eta p}) t_{\eta p \to \eta p} (M_{\eta p}) \nonumber \\
&& + \frac{\sqrt{2}}{2} G_{\pi^0 p} (M_{\eta p}) t_{\pi^0 p \to \eta p} (M_{\eta p}) \nonumber \\
&& + G_{\pi^+ n} (M_{\eta p}) t_{\pi^+ n \to \eta p} (M_{\eta p}) \nonumber \\
&& - \frac{\sqrt{6}}{3} G_{K^+ \Lambda} (M_{\eta p}) t_{K^+ \Lambda
\to \eta p} (M_{\eta p}) \Big),   \label{Eq:tmb}
\end{eqnarray}
where $V_P$ expresses the weak and hadronization strength, which is
assumed to be a constant and independent of the final state
interaction. In the above equation, $G_{MB}$ denotes the
one-meson-one-baryon loop function, which depends on the invariant
mass of the final $\eta p$ system, $M_{\eta p}$. The meson-baryon
scattering amplitudes $t_{MB \to \eta p}$ are those obtained in the
chiral unitary approach, which depend also on $M_{\eta p}$. Details
can be found in Refs.~\cite{Inoue:2001ip,Doring:2005bx}.

\subsection{Effective Lagrangian approach and  the $N^*(1535)$ resonance as a Breit-Wigner resonance}

\begin{figure}[htbp]
\begin{center}
\includegraphics[scale=0.8]{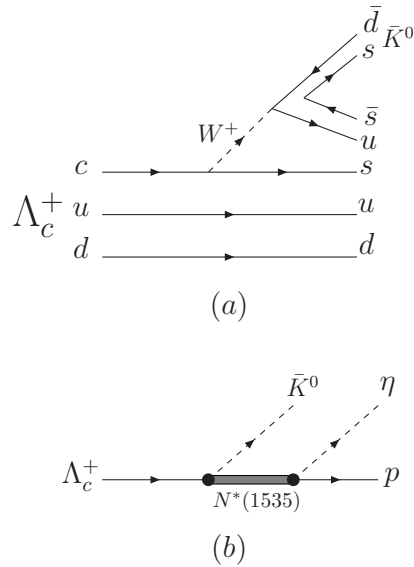}
\caption{Quark level diagram for $\Lambda^+_c \to
\bar{K}^0 N^*(1535)$ (a) and hadron level diagram for $\Lambda^+_c
\to \bar{K}^0 \eta p$ decay (b).} \label{Fig:Nstar1535}
\end{center}
\end{figure}

On the other hand, because the $N^*(1535)$ has a large $uud s \bar s$
component, it can also be produced via the process shown in
Fig.~\ref{Fig:Nstar1535} (a), similar to the $P^+_c$ states produced
in the $\Lambda^0_b \to K^- P^+_c$ decay~\cite{Aaij:2015tga}. After
the $N^*(1535)$ is formed with $uud s \bar s$, it decays into 
$\eta p$, which is the dominant decay channel of the $N^*(1535)$
resonance. We show the hadron level diagram for the decay of
$\Lambda^+_c \to \bar{K}^0 N^*(1535) \to \bar{K}^0 \eta p$ in
Fig.~\ref{Fig:Nstar1535} (b).

Before going further, we emphasize that the strangeness component of $N^*(1535)$ can not be guaranteed from the decay process shown in Fig.~\ref{Fig:Nstar1535}. Indeed, the $N^*(1535)$ can also be produced from the process shown in Fig.~\ref{Fig:mbproduction}, where the $s\bar{d}$ forms the $\bar{K}^0$, while the $N^*(1535)$ is constructed from the $uud$ cluster and then it decays into $\eta p$ because of its large coupling to this channel. 

The general decay amplitudes for $\Lambda^+_c \to \bar{K}^0
N^*(1535)$ can be decomposed into two different structures with the
corresponding coefficients $A$ and $B$,
\begin{eqnarray}
{\cal M} = i \bar{u}(q) (A+B\gamma_5)u(p),
\label{Eq:LambdactoKbarNstar}
\end{eqnarray}
where $q$ and $p$ are the momentum of the $N^*(1535)$ and $\Lambda^+_c$,
respectively. The coefficients $A$ and $B$ for charmed baryons decaying into ground meson and baryon states, in
general, can be calculated in the framework of the pole
model~\cite{Cheng:1991sn} or within the perturbative $QCD$
approach~\cite{Lu:2009cm}. In the present case, because the
$N^*(1535)$ resonance is not well understood in the classical quark
model, the values of $A$ and $B$ in
Eq.~\eqref{Eq:LambdactoKbarNstar} are very difficult to be pined down,
and we have to determine them with future  experimental
data. In this work, we take $A = B$ and we come back to this issue later.

To get the whole decay amplitude of $\Lambda^+_c \to \bar{K}^0
N^*(1535) \to \bar{K}^0 \eta p$ as shown in Fig.~\ref{Fig:Nstar1535}
(b), we use the effective Lagrangian density of
Refs.~\cite{Bai:2001ua,Liu:2005pm,Xie:2007qt} for the $N^*(1535)N\eta$
vertex,
\begin{eqnarray}
{\cal L}_{N^*N\eta} = -i g_{N^*N\eta} \bar{N} \eta N^* + {\rm h.c.},
\end{eqnarray}
where $N$, $\eta$, and $N^*$ represent the fields of the proton, the $\eta$
meson, and the $N^*(1535)$ resonance, respectively.

The invariant decay amplitude of the $\Lambda^+_c \to \bar{K}^0
N^*(1535) \to \bar{K}^0 \eta p$ decay is
\begin{eqnarray}
T^{N^*} = i
g_{N^*N\eta}\bar{u}(p_3,s_p)G_{N^*}(q)(A+B\gamma_5)u(p,s_{\Lambda^+_c}),
\end{eqnarray}
where $p_3$ is the momentum of the final proton. The $s_p$ and
$s_{\Lambda^+_c}$ are the spin polarization variables for the proton
and $\Lambda^+_c$ baryon, respectively. The $G_{N^*}(q)$ is the
propagator of the $N^*(1535)$, which is given by a  Breit-Wigner (BW)
form as,
\begin{eqnarray}
G_{N^*} (q) = i \frac{\Slash q + M_{N^*}}{q^2 - M^2_{N^*} + i
M_{N^*} \Gamma_{N^*}(q^2)},
\end{eqnarray}
where $M_{N^*}$ and $\Gamma_{N^*}(q^2)$ are the mass and total decay
width of the $N^*(1535)$, respectively. We take $M_{N^*} =
1535$ MeV as in the PDG~\cite{Olive:2016xmw}. For
$\Gamma_{N^*}(q^2)$, since the dominant decay channels for the
$N^*(1535)$ resonance are $\pi N$ and $\eta N$~\cite{Olive:2016xmw},
we take the following form as used in
Refs.~\cite{Wu:2009nw,Xie:2013wfa}
\begin{eqnarray}
\Gamma_{N^*} (q^2) = \Gamma_{N^* \to \pi N}(q^2) + \Gamma_{N^* \to
\eta N}(q^2) + \Gamma_0,   \label{Eq:GamrNstarq2}
\end{eqnarray}
with
\begin{eqnarray}
\Gamma_{N^* \to \pi N}(q^2) \! &=& \!\! \frac{3g^2_{N^*N\pi}}{4\pi} \frac{\sqrt{|\vec{p}_{N\pi}| + m^2_p} + m_p}{\sqrt{q^2}} |\vec{p}_{N\pi}|, \\
\Gamma_{N^* \to \eta N}(q^2) \! &=& \!\! \frac{g^2_{N^*N\eta}}{4\pi}
\frac{\sqrt{|\vec{p}_{N\eta}| + m^2_p} + m_p}{\sqrt{q^2}}
|\vec{p}_{N\eta}|.
\end{eqnarray}
Here
\begin{eqnarray}
|\vec{p}_{N\pi}| &=&
\frac{\lambda^{1/2}(q^2,m^2_p,m^2_{\pi})}{2\sqrt{q^2}}, \\
|\vec{p}_{N\eta}| &=&
\frac{\lambda^{1/2}(q^2,m^2_p,m^2_{\eta})}{2\sqrt{q^2}},
\end{eqnarray}
where $\lambda$ is the K\"allen function with $\lambda(x,y,z) =
(x-y-z)^2 -4yz$. In the present work, we take $g^2_{N^*N\pi}/4\pi = 0.037$ and
$g^2_{N^*N\eta}/4\pi = 0.28$ as used in Ref.~\cite{Lu:2014yba}. With
these values we can get $\Gamma_{N^* \to N\pi} = 67.5$ MeV and
$\Gamma_{N^* \to N\eta} = 63$ MeV if we take $\sqrt{q^2} = 1535 $
MeV. In this work, we choose $\Gamma_0 = 19.5$ MeV for
$\Gamma_{N^*}(\sqrt{q^2} = 1535 ~{\rm MeV}) = 150$ MeV.

In the effective Lagrangian approach, the sum over polarizations and
the Dirac spinors can be easily done thanks to
\begin{eqnarray}
\sum_{s_p} \bar{u}(p_3,s_p) u(p_3,s_p) &=& \frac{\Slash p_3 + m_p}{2
m_p}, \\
\sum_{s_{\Lambda^+_c}} \bar{u}(p,s_{\Lambda^+_c})
u(p,s_{\Lambda^+_c}) &=& \frac{\Slash p + M_{\Lambda^+_c}}{2
M_{\Lambda^+_c}}.
\end{eqnarray}
Finally, we obtain
\begin{eqnarray}
{1\over 2} \overline{\sum_{s_{\Lambda^+_c}}} \sum_{s_p}{|T^{N^*}|}^2
&=& \frac{g^2_{N^* N \eta}}{2 m_p
M_{\Lambda^+_c} |D|^2} \times \nonumber \\
&& \left [ (ap\cdot q + bp_3 \cdot p + cM_{\Lambda^+_c})A^2 +
\right.
\nonumber \\
&& \left. (ap\cdot q + bp_3 \cdot p - cM_{\Lambda^+_c})B^2 \right ],
\label{amplitudesquare-Nstar1535}
\end{eqnarray}
with
\begin{eqnarray}
D &=& q^2 - M^2_{N^*} + i M_{N^*} \Gamma_{N^*}(q^2), \\
a &=& 2(p_3 \cdot q + m_pM_{N^*}), \\
b &=& M^2_{N^*} - q^2 ,\\
c &=& m_p(M^2_{N^*} + q^2) + 2M_{N^*}p_3 \cdot q,
\end{eqnarray}
and
\begin{eqnarray}
p \cdot q &=& \frac{M^2_{\Lambda^+_c} + M^2_{\eta p} -
m^2_{\bar{K}^0}}{2}, \\
p_3 \cdot q &=& \frac{M^2_{\eta p} + m^2_p - m^2_{\eta}}{2}, \\
p_3 \cdot p \! \! &=& \!\! \frac{(M^2_{\Lambda^+_c} + M^2_{\eta p} -
m^2_{\bar{K}^0})(M^2_{\eta p} + m^2_p - m^2_{\eta})}{2M^2_{\eta p}},
\end{eqnarray}
with $M^2_{\eta p} = q^2$.

\subsection{Invariant mass distributions of the $\Lambda^+_c \to \bar{K}^0 \eta p$ decay}

With all the ingredients obtained in the previous subsection, one
can write down the invariant $\eta p$ mass distribution of the $\Lambda^+_c
\to \bar{K}^0 \eta p$ decay as:
\begin{eqnarray}
\frac{d \Gamma}{d M_{\eta p}} &=& \frac{1}{16\pi^3} \frac{m_p
p_{\bar{K}^0} p^*_{\eta}}{M_{\Lambda^+_c}} \left\lvert T
\right\rvert^2 , \label{Eq:dgdmdm}
\end{eqnarray}
where $T$ is the total decay amplitude. The $p_{\bar{K}^0}$ and $p^*_{\eta}$ are the
three-momenta of the outgoing $\bar{K}^0$ meson in the
$\Lambda^+_c$ rest frame and the outgoing $\eta$ meson in the center
of mass frame of the final $\eta p$ system, respectively. They are
given by
\begin{eqnarray}
p_{\bar{K}^0} &=& \frac{\lambda^{1/2}(M^2_{\Lambda^+_c},M^2_{\eta p},m^2_{\bar{K}^0})}{2M_{\Lambda^+_c}}, \\
p^*_{\eta} &=& \frac{\lambda^{1/2}(M^2_{\eta
p},m^2_\eta,m^2_p)}{2M_{\eta p}}.
\end{eqnarray}

The range of $M_{\eta p}$ is
\begin{eqnarray}
M^{\rm max}_{\eta p} & = & M_{\Lambda^+_c} - m_{\bar{K}^0},
\nonumber \\
M^{\rm min}_{\eta p} &=& m_{\eta} + m_p . \nonumber
\end{eqnarray}

\section{Numerical results and discussion} \label{Sec:Results}

In this section, we first show the numerical results for the $d
\Gamma/d M_{\eta p}$ with three models: Model I takes $T= T ^{MB}$;
Model II takes  $T= T ^{N^*}$ and $\Gamma_{N^*}$ is energy dependent
as in Eq.~\eqref{Eq:GamrNstarq2}; Model III takes $T= T ^{N^*}$ and
$\Gamma_{N^*} = 150$ MeV as a constant. Next, we will discuss the
impact of the contributions from other $N^*$ and $\Sigma^*$
states.

\subsection{Invariant $\eta p$ mass distributions}

\begin{table}[htbp]
\begin{center}

\caption{\label{tabmass} Masses and spin-parities of the particles
studied in the present work.}

\begin{tabular}{ccc}

\hline\hline

State          & Mass (MeV)  &  Spin-parity ($J^P$)    \\

\hline

$\Lambda^+_c$    & $2286.46$   & $\frac{1}{2}^+$ \\
$\bar{K}^0$          & $497.61$    & $0^-$           \\

$\eta$           & $547.86$    & $0^-$            \\

$p$        & $938.27$   & $\frac{1}{2}^+$  \\

\hline\hline

\end{tabular}
\end{center}

\end{table}

\begin{figure}[htbp]
\begin{center}
\includegraphics[scale=0.45]{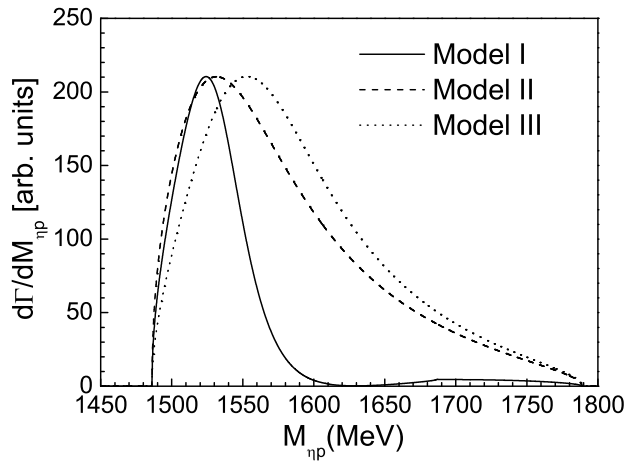}
\caption{Invariant $\eta p$ mass distribution for the $\Lambda^+_c
\to \bar{K}^0 \eta p$ decay. The solid, dashed, and dotted curves
represent the results obtained in Model I, II, and III,
respectively.} \label{Fig:dgdm-etap}
\end{center}
\end{figure}

In Fig.~\ref{Fig:dgdm-etap}, we show the $\eta p$ invariant mass
distribution obtained with the mass values shown in
Table~\ref{tabmass}, where the solid, dashed and dotted curves
represent the numerical results obtained with Model I, II, and III,
respectively. The results of Model I are obtained with $V_P = 1$
MeV$^{-1}$. The results of Model II with $A=B=45.2$ and Model III
with $A=B=47.4$ are normalized to the peak of Model I.

For Model I, a peak around $1524$ MeV corresponding to the
$N^*(1535)$ resonance can be clearly seen as in
Refs.~\cite{Inoue:2001ip,Doring:2005bx}. The peaks of Model II and
III move to $1532$ and $1553$ MeV, respectively. The peak position
of Model II is very close to the central value, 1535 MeV, estimated
in the PDG~\cite{Olive:2016xmw} for the $N^*(1535)$. The peak
position of Model I is also close to the value 1535 MeV, but with a
narrow width. For Model III where a constant decay width of
the $N^*(1535)$ is used, the peak position moves $20$ MeV away from the
Breit-Wigner mass $1535$ MeV. Besides, the resonant shapes of Model
II and III are broader than the result of Model I.

Because the mass of the $N^*(1535)$ is close to the $\eta N$
threshold and has a  large coupling to this channel, the approximation
of a BW form with a constant width is not very realistic~\cite{Liu:2005pm}. We
should take the coupled channel BW formula as in
Eq.~\eqref{Eq:GamrNstarq2}, which will reduce the BW mass of the
$N^*(1535)$~\cite{Liu:2005pm}.

From the results of Model I and II shown in
Fig.~\ref{Fig:dgdm-etap}, we see that these two different
descriptions of the $N^*(1535)$ resonance give different invariant
$\eta p$ mass distributions. The findings here are similar to that
obtained in Refs.~\cite{Doring:2008sv,Geng:2008cv}. For the
$N^*(1535)$, the amplitude square obtained with the chiral
unitary approach does not behave like an usual BW resonance, even at
the peak position (see Fig. 1 of Ref.~\cite{Geng:2008cv}). It is
expected that future experimental measurements may test our
model predictions and clarify this issue.

One might be tempted to think that the discrepancy between Model I and II (or III)
is due to the inclusion of the $p$-wave contribution for Model II
and III shown in Eq.~\eqref{Eq:LambdactoKbarNstar} with the $B$ term. We
have explored such a possibility from the comparison of the
contributions of the $A$ and $B$ terms. For doing this, we first rewrite
$d\Gamma/dM_{\eta p}$ for Model II and III as,
\begin{eqnarray}
\frac{d \Gamma}{d M_{\eta p}} &=& f_1 A^2 + f_2 B^2.
\end{eqnarray}
Then we define the ratio $R$ as
\begin{eqnarray}
R = \frac{f_2 B^2}{f_1 A^2} = \frac{f_2}{f_1}.
\end{eqnarray}
In the last step, we have taken $A = B$.

\begin{figure}[htbp]
\begin{center}
\includegraphics[scale=0.45]{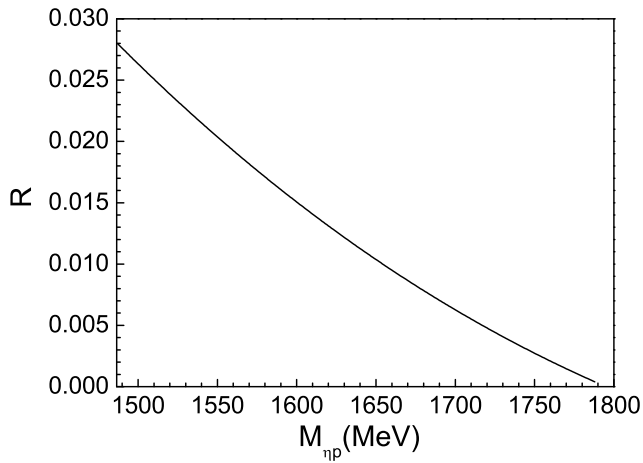}
\caption{Ratio $R$ of the $B$ and $A$ terms as a function of the
$\eta p$ invariant mass.} \label{Fig:ratio}
\end{center}
\end{figure}

In Fig.~\ref{Fig:ratio} we show the numerical results for $R$ as a
function of $M_{\eta p}$. We see clearly that $R$ is less than $2.8$
percent for the whole possible $M_{\eta p}$ in the $\Lambda^+_c \to
\bar{K}^0 \eta p$ decay. This means that the contribution of the
$p$-wave $B$ term is rather small in comparison with the contribution from the
$s$-wave $A$ term and can be neglected safely. This study provides
further support for the factorization scheme of the hard process
(the weak decay and hadronization) for Model I where only the $s$-wave
contribution is considered between any  two particles of the final 
$\bar{K}^0 \eta p$. Such a factorization scheme seems to work fairly
well in the present case.

It should be noted that the $B$ term is very small compared with the $A$
term, which is tied to the fact that we take $A = B$. A model
independent calculation of the values of $A$ and $B$ is most welcome
and will ultimately test our model calculations.

\subsection{Contributions  from other processes}

Up to now, we have considered only the contribution from $N^*(1535)$, while the contributions from other nucleon resonances,
such as, $N^*(1650)\frac{1}{2}^-$, $N^*(1710)\frac{1}{2}^+$, and
$N^*(1720)\frac{3}{2}^+$, are not taken into account. The
$N^*(1710)$ and $N^*(1720)$ decay into $\eta p$ in $p$-wave and the
decay of $\Lambda^+_c \to \bar{K}^0 N^*(1710)$ and $\Lambda^+_c \to
\bar{K}^0 N^*(1720)$ have very limited phase space, hence, their
contributions should be much suppressed.

It is interesting to note that both $N^*(1535)$ and $N^*(1650)$
are dynamically generated from the analysis of the $s$-wave $\pi N$
scattering \cite{Nieves:2001wt,Bruns:2010sv}. We list the results
obtained in
Refs.~\cite{Nieves:2001wt,Bruns:2010sv,Nieves:2011gb,Khemchandani:2013nma,Garzon:2014ida}
for $N^*(1535)$ and $N^*(1650)$ in Table~\ref{tabNstar}, where we
show also the BW mass and width, branching ratios to $\pi N$ and
$\eta N$ of $N^*(1535)$ and $N^*(1650)$ that are estimated by
PDG~\cite{Olive:2016xmw} for comparison. We see that the 
$N^*(1650)$  and the $N^*(1535)$ are much separated in mass~\cite{Nieves:2001wt,Bruns:2010sv}. Therefore, the contribution from $N^*(1650)$
will not overlap too much with that from $N^*(1535)$. Furthermore, the
branching ratio of $N^*(1650)$ to $\eta N $ is very small compared with
the one to $\pi N$. We thus conclude that the contribution from
$N^*(1650)$ to the invariant $\eta p$ mass distribution is small or at least it will not change too much the numerical results shown in
Fig.~\ref{Fig:dgdm-etap} even if it has a sizable contribution.

\begin{table}[htbp]
\begin{center}

\caption{\label{tabNstar} Mass ($M_R$) and width ($\Gamma_R$) for
$N^*(1535)$ and $N^*(1650)$ found in
Refs.~\cite{Nieves:2001wt,Bruns:2010sv,Nieves:2011gb,Khemchandani:2013nma,Garzon:2014ida}.
The masses and widths from Ref.~\cite{Olive:2016xmw} are deduced
from a Breit-Wigner fit. The values of masses and widths are given
in MeV.}

\begin{tabular}{c|cc|cc}

\hline\hline

Reference         & \multicolumn{2}{c|}{$N^*(1535)$}    & \multicolumn{2}{c}{$N^*(1650)$}   \\
& $M_R$ & $\Gamma_R$  &  $M_R$ & $\Gamma_R$    \\

\hline

\cite{Nieves:2001wt}    & $1496.5 \pm 0.4$ & $83.3 \pm 0.7$   & $1684.3 \pm 0.7$ & $194.3 \pm 0.8$ \\
\cite{Bruns:2010sv}          & $1506$ & $280$    & $1692$ & $92$           \\
\cite{Nieves:2011gb,Gamermann:2011mq}     & $1556$ & $94$    & $1639$ & $76$     \\
\cite{Khemchandani:2013nma}     & $1504$ & $110$    & $1668$ & $56$     \\
                                &        &          & $1673$ & $134$~\footnote{A twin pole structure for $N^*(1650)$ is obtained  in Ref.~\cite{Arndt:1995bj}.}     \\
\cite{Garzon:2014ida}     & $1508.1$ & $90.3$    & $1672.3$ & $158.2$     \\
\cite{Olive:2016xmw}           & $1535 \pm 10$    & $150 \pm 25$  & $1655 \pm 15$ & $140 \pm 30$          \\
 \hline
${\rm Br}(\to \pi N)$        & \multicolumn{2}{c|}{$35 \sim 55 \%$}   & \multicolumn{2}{c}{$50 \sim 70 \%$} \\
${\rm Br}(\to \eta N)$        & \multicolumn{2}{c|}{$32 \sim 52 \%$}
& \multicolumn{2}{c}{$14 \sim 22 \%$~\footnote{This value is quoted
in PDG \cite{Olive:2016xmw}, but it is originally taken from
Ref.~\cite{Anisovich:2011fc} which is derived from a multichannel
partial wave analysis of pion and photo-induced reactions off
protons. On the other hand, from the coupled-channel analysis of
$\eta$ meson production including all recent photo-production data
on the proton, the value of $1 \pm 2 \%$ is obtained in Ref.~\cite{Shklyar:2012js} and of $1.4\%$ in Ref.~\cite{Shklyar:2004ba}.}} \\
\hline\hline

\end{tabular}
\end{center}

\end{table}

\begin{figure}[htbp]
\begin{center}
\includegraphics[scale=0.45]{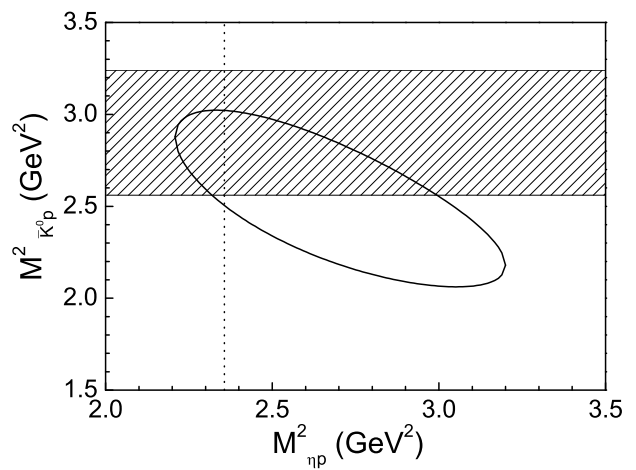}
\caption{Dalitz plot for $M^2_{\eta p}$ and $M^2_{\bar{K}^0 p}$ in
the $\Lambda^+_c \to \bar{K}^0 \eta p$ decay. The $N^*(1535)$ energy
is shown by the vertical dotted line, which the horizontal band
represents the masses of $\Sigma^*$ states from 1600 to 1800 MeV.}
\label{Fig:dalitzplot}
\end{center}
\end{figure}

On the other hand, there should be also contributions from
$\Sigma^*$ resonances that have significant branching ratio to
$\bar{K}^0 p$. Those $\Sigma^*$ resonances are:
$\Sigma^*(1660)\frac{1}{2}^+$, $\Sigma^*(1670)\frac{3}{2}^-$, and
$\Sigma^*(1750)\frac{1}{2}^-$. We show the Dalitz plot for the
$\Lambda^+_c \to \bar{K}^0 \eta p$ decay in
Fig.~\ref{Fig:dalitzplot}. In the $N^*(1535)$ energy region, the
Dalitz plot overlaps  with these $\Sigma^*$ resonances from
$1600$ to $1800$ MeV in the $\bar{K}^0 p$ channel, which may make
the analysis of $N^*(1535)$ difficult. Fortunately, the
$\Sigma^*(1660)\frac{1}{2}^+$ and $\Sigma^*(1670)\frac{3}{2}^-$
decay into $\bar{K}^0 p$ in $p$-wave and $D$-wave, respectively.
These contributions will be suppressed because of the higher partial
waves involved. For the $\Sigma^*(1750)\frac{1}{2}^-$, it decays into $\bar{K}^0
p$ in $s$-wave. However, it lies in the kinematic end-point region and therefore the
decay of $\Lambda^+_c \to \eta \Sigma^*(1750)$ has a relative small
phase space.

In summary, the contributions from other $N^*$ and $\Sigma^*$
resonances should be small compared with the contribution from the
$N^*(1535)$, and we expect that their contributions
will not change much the model predictions presented in the present work. On the other hand,  if 
future experimental measurements provide enough data to disentangle the contributions from these resonances, one can also study them.
It should be kept in mind that our study made some assumptions and hence it can be improved once more data become available.

\section{Conclusions}

In the present work we have studied the invariant $\eta p$ mass
distribution in the $\Lambda^+_c \to \bar{K}^0 \eta p$ decay to better
understand the $N^*(1535)$ resonance. First, we employed the  molecular picture where the $N^*(1535)$ is
dynamically generated from the meson-baryon interaction. In such a scenario, the weak
interaction part is dominated by the $c$ quark decay process: $c(ud)
\to (s + u + \bar d)(ud)$, while the hadronization part takes place
by the $uud$ cluster picking up a $q\bar{q}$ pair from the vacuum and
hadronizes into a meson-baryon pair, while the $s \bar{d}$ pair from
the weak decay turns into a $\bar{K}^0$.  The following final state
interactions of the meson-baryon pairs are described in the chiral
unitary model that dynamically generates the $N^*(1535)$ resonance
in the $I = 1/2$ sector. Second, we  studied the $\Lambda^+_c \to
\bar{K}^0 N^*(1535) \to \bar{K}^0 \eta p$ decay with a Breit-Wigner
formula to describe the distribution of the $N^*(1535)$ in the
effective Lagrangian model. The above two descriptions for
the $N^*(1535)$ resonance give different invariant $\eta p$ mass
distributions. Furthermore, we showed in a qualitative way that the contributions from other $N^*$ and $\Sigma^*$ resonance are relatively small and
 will not affect much the results obtained in the present study.

On the experimental side, the decay mode $\Lambda^+_c \to \bar{K}^0
\eta p$ has been observed~\cite{Olive:2016xmw,Ammar:1995je} and the
branching ratio $\mathrm{Br}(\Lambda^+_c \to \bar{K}^0 \eta p)$ is
determined to be $(1.6 \pm 0.4)\%$, which is one of the dominant
decay modes of the $\Lambda^+_c$ state. For the decay of
$\Lambda^+_c \to \bar{K}^0 \eta p$, the final $\eta p$ is in pure
isospin $I = 1/2$. Hence, this decay can be an ideal process to
study the $N^*(1535)$ resonance, which has a large branching ratio
to $\eta N$ and decays into $\eta N$ in $s$-wave. Future
experimental measurements of the invariant $\eta p$ mass
distribution studied in the present work will be very helpful to
test our model calculations and constrain the properties of
the $N^*(1535)$ resonance. For example, a corresponding experimental
measurement could in principle be done at
BESIII~\cite{Ablikim:2015flg} and Belle.

\section{ACKNOWLEDGMENTS}

This work is partly supported by the National Natural Science
Foundation of China under Grant No.11475227, No. 11375024, No. 11522539, No. 11505158, No. 11475015, and No. 11647601. It is also supported by the Youth Innovation Promotion
Association CAS (No. 2016367).

\bibliographystyle{ursrt}

%\end{spacing}
\end{document}